# Proposal of an experimental strategy to introduce Relativity in elementary physics courses

*Roberto Assumpção*[1,2]
1. PUC–Minas, Faculty of Electrical Engineering and Telecommunication
Av. Pe. Francis C Cox, Poços de Caldas, MG, Brasil, 37701-355 assumpcao@pucpcaldas.br
2. UNICAMP – Faculty of Mechanical Engineering, Cidade Universitária "Zeferino Vaz",
Campinas, SP, Brasil, 13083-970 assump@fem.unicamp.br

ABSTRACT – This contribution shows that the main topics of *quantum and relativity* theories can be discussed at an elementary level and in a considerable extent, including the formal results of the "Uncertainty Relations", "Time Dilation" and "Lorentz Contraction" , by a minor modification of the usual mathematical formalism employed to describe uniform and accelerated motion. The proposed strategy begins with a "free–fall" experiment, followed by a discussion of the results in order to include deviations of the experimental data with respect to the conventional kinematics model.

## INTRODUCTION

Introducing the basic principles of Mechanics, as early as possible, is a desire of physical educators, even at the sacrifice of a rigorous mathematical treatment, provided the concepts can be preserved. Nowadays, first–year university students of physical sciences have such a spectrum of information concerning Gravitation, Quantum theory, Relativity and applications, particularly in the domain of Astrophysics, that the 'desire' transforms into a real necessity.

This fact rests, probably, not only on information; we have to recognise that the evolution of physics led to, say, four mechanics: Classical, Quantum, Relativistic and Statistical mechanics. In effect, the structure of formal and non–formal education is such that secondary school students have knowledge of the first and at least good information concerning Quantum and Relativistic theories.

On this basis, therefore, it is difficult to explain just the *Laws of Newton* to students incoming to the university; frequently asked questions range from the general validity of the laws to details such as the Uncertainty Principle, "twin paradox", and the behaviour of a variety of objects, such as photons, electrons, satellites, etc. .

We are thus lead to the task of teaching Classical Mechanics, while preserving the curiosity of the students on those up-to-date topics; during the last semesters we started an attempt to extract the main concepts from Einstein's formulation and present them in an introductory undergraduate physics course directed to Electrical Engineering students. In our view, the experimental way has proved to be quite precise, and that is the procedure here described.

## EXPERIMENTAL PROCEDURE AND ANALYSIS

Relativity and also Quantum Mechanics [1] are usually introduced via historical aspects paralleling a critical review of some crucial experiments; however, this approach is out of the scope in introductory courses [2], once the students have no background other than secondary school physics.

The procedure employed here starts with the experimental determination of 'g' – the acceleration of gravity: about 20 students, divided in groups of 4, are asked to measure the height and time of fall of three blocks of distinct masses and determine 'g', by plotting the corresponding height against the medium value of the elapsed time, H (t). A technical report of this experiment should be presented after week.

The relatively low frequency of good results ( g-values may typically range from 5 m/s$^2$ to 15 m/s$^2$ ! ) causes embarrassment. Most students reject the experimental data and some even refuse to present a report.

Then a laboratory experiment was carried out, employing passage photocell detectors connected to a chronometer (Fig. 1).

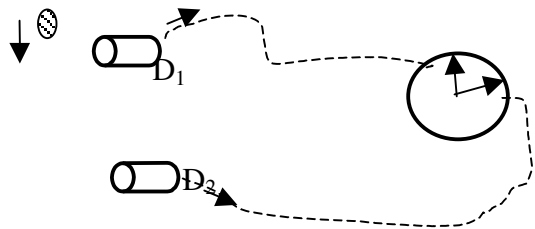

**Figure 1**. Free– Fall Laboratory Experiment

Here the first detector turns the chronometer ON, while the second turns it OFF; by varying the distance – 'H' between detectors and registering the time of fall, a plot of H(t) leads to the value of the local gravity.

Students are asked to plot H x t and also H x t$^2$ ; Figure 2 shows a typical result.

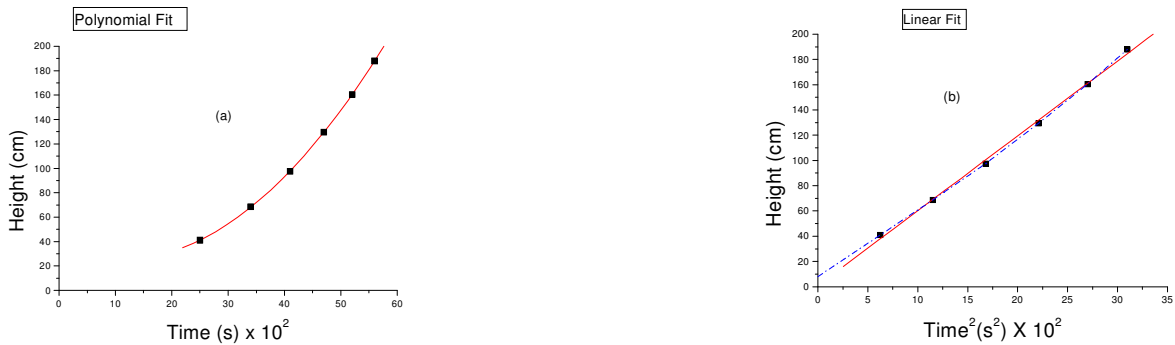

**Figure 2** (a) Parabolic – H x t and (b) linear – H x t$^2$ plots of a "free-fall experiment;
The dashed curve in (b) is a numerical extrapolation including all data. points.

This second procedure gives a more reliable ( about 10 m/s$^2$ ) **g**–value; however, comparing the experimental curves with the theoretical prediction, we note the presence of a small Y-axis component. The familiar equation of motion is:

$$H(t) = H(o) + V(o)t + \frac{1}{2}gt^2 \quad (1)$$

For a body abandoned quite near the first detector, the relation should be just

$$H(t) = \frac{1}{2}gt^2 \quad (2)$$

Thus, the existence of an initial Y-component implies that either the initial velocity or the initial height (namely both) are distinct from zero :

$$\Delta Y = H(o) \quad \text{and/or} \quad \Delta Y = V(o)t \quad (3)$$

At this point a quantum – mechanical discussion of *Uncertaint* could be done; however, we postpone this in order to centre the problem on Time. The reason is that – at this stage – students are aware of the importance of a precise time measurement in order to obtain a good **g**–value, since the first experiment was carried out "by their own", employing conventional clocks and visually detecting the motion, whereas the second was a concrete laboratory test that includes automatic detection.

Thus we consider the existence of a "Hidden Time" – $t^*$, or a time not detected by the instrumentation, which is responsible for the initial $\Delta Y$ in terms of initial velocity X time ( $V_o$ x $t^*$ ) . Due to equations (3), students can easily comprehend $t^*$ also as a means of expressing $\Delta Y$ as a function of an initial ' imprecision' of height ($\delta H$)

In order to increase the experimental time with respect to the "Hidden Time" we are naturally directed to analyse the motion of a brick sliding down an incline. This was the procedure adopted by Newton [3] and also Galileo [4] to study acceleration.

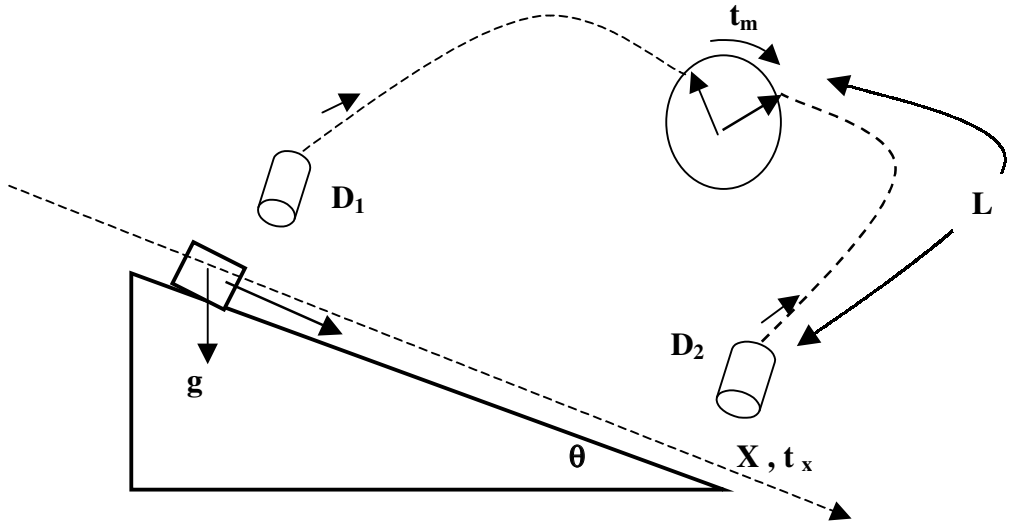

**Figure 3.** A Brick sliding down an incline. $D_1$ and $D_2$ are photo-detectors; $t_m$ is the time registered by the chronometer a distance '**L**' apart from the plane.

Now we focus attention on description of the motion; according to Figure 3, there are three times involved:

$t_x$ , which is the effective time of motion, or the time present as a parameter in the equation of motion (1).
$t_m$ , the measured time of motion, or the time detected by the chronometer.
$t^*$ , the "Hidden Time" , which is related to the time required to start the detection of motion.

Students are asked to determine **g** from the experimental data but mainly to write the equations of motion in terms of $t_m$ and $t_x$ . At this point they easily realise that the plots of Figure 2 are, in fact, H = H ($t_m$) , that is, an experimental relation; conversely, equation (1) is H = H ($t_x$ ), that is, a mathematical ( theoretical) model . The connection between the experimental and effective time is:

$$t_m = t_x + L/v_s \quad (4)$$

$v_s$ being the signal velocity trough the wire of length **L**.

The "Hidden Time" $t^*$ appears explicitly as a time (delay) required to transport the information from the plane to the chronometer:

$$t^* \equiv t_i = L/v_s,  \quad (5)$$

The component of acceleration along the incline is $a = g\,sen\,\theta$; thus if the brick is abandoned from rest – $v_{ox} = 0$, we have the familiar equation: $v_x = g\,sen\,\theta\,t_x$.

However, the experimental arrangement tells that when the chronometer starts to run, the brick should have a component of velocity given by:

$$v_o = g\,sen\,\theta\,L/v_s \quad (6)$$

Note that this is a Classical or a Causality concept, that is: if one event ( the motion of the brick ) is the cause of a second ( the motion of the chronometer ), then the second event must surely succeed the first, in any coordinate system. Indeed, we can write the kinetic equations in the chronometer ( $t_m$ ) coordinate system:

$$x(t_m) = g\,sen\,\theta\,\frac{L}{v_s}t_m + \frac{g\,sen\,\theta\,t_m^2}{2} \quad (7)$$

Conversely, the equation in the incline ($t_x$) coordinate system is:

$$x(t_x) = \frac{g\,sen\,\theta\,t_x^2}{2} \quad (8)$$

But the time $t_x$ in equation (8) corresponds to a clock simultaneous with the brick; in relativistic terms, this clock registers the *proper time* of the brick. At this point students realise that **$t_x$** is obtained just by means of **$t_m$**; moreover, they arrive to the theoretical aspect of **$t_x$**. Thus we believe that this is the precise moment to discuss Simultaneity in terms of flux of information; taking relation (4), we have:

$$t_m = t_x + \frac{L}{v_s} \quad (9)$$

Dividing by $t_m$ and rearranging,

$$\frac{t_x}{t_m} = 1 - \frac{L}{t_m v_s} = (1-\alpha) \quad (10)$$

$$\alpha \equiv \frac{L}{t_m v_s} = \frac{t_i}{t_m}$$

Now relation (10) is analogous to the relativistic expression for the *Time Dilation* effect:

$$t_x = (1-\alpha)t_m \quad (11)$$

This way, the concept of *Proper Time* is introduced as a necessary way to express the results of experimental data; moreover, students can understand *Time Dilation* as a natural effect that appears in Relativity theory and not in newtonian mechanics, since the factor $\alpha$ tends to zero in the classical limit ( $t_i << t_m$ ).

The interesting feature of this formula is that it is independent of the notion of moving frames; in other words, we arrived at (11) in order to avoid the usual Relativity discussion of frames in an introductory course, when the students are just learning how scale, plot and analyse fixed frames.

In our view, this approach puts Relativity into a sounded basis, especially for Electrical Engineering students; discussion of triggering, response time and circuitry impedance come naturally as a consequence of the *Time Dilation* effect.
Another qualitative aspect is that students can work out (11) comparing it with the equations shown in standard textbooks [5,6]; suppose, for instance, that ' L' is made parallel to the direction X in Fig 3; then the fraction $L/t_m$ represents the velocity of the brick, say, $v_m$. Further, suppose a low index optical fibber transporting the signal, that is, $v_s \sim c$. Then, equation (11) becomes:

$$t_x \approx (1-\frac{v_m}{c})t_m \quad (11.a)$$

revealing an extremely formal equivalence with the familiar relativistic expression.

A discussion of Lorentz contraction is also possible if the teacher takes the necessary care on dealing with the concepts of space and length; from equations (7) and (8) we have:

$$\frac{x(t_m)}{x(t_x)} = \frac{g \sen\theta \frac{L}{v_s} t_m + \frac{1}{2} g \sen\theta\, t_m^2}{\frac{1}{2} g \sen\theta\, t_x^2} \quad (12)$$

By means of (9) we have:

$$\frac{x(t_m)}{x(t_x)} = \frac{t_m^2 + 2\frac{L}{v_s} t_m}{t_m^2 - 2\frac{L}{v_s} t_m + \frac{L^2}{v_s^2}} \quad (13)$$

Dividing by $t_m^2$ and neglecting the second order term in the denominator,

$$\frac{x(t_m)}{x(t_x)} \cong \frac{(1 + 2\frac{L}{v_s t_m})}{(1 - 2\frac{L}{v_s t_m})} = \frac{(1+2\alpha)}{(1-2\alpha)} \quad (14)$$

rearranging,

$$x(t_x) = \frac{(1-2\alpha)}{(1+2\alpha)} x(t_m) \quad (15)$$

which can be written as,

$$x(t_x) \approx (1 - 4\alpha) x(t_m) \quad (16)$$

where we used the binomial theorem retaining only the first order term.

We then have a peculiar result; this formulae is reasonable to describe the Lorentz contraction, or the contraction in the length of a rigid body moving in a direction parallel to that length, though this is not *Length Contraction*.

But this would be consistent with our approach here if historical [7] aspects are inserted on the discussion; the reason is that Lorentz effectively used his formulation in order to explain the null result of the Michelson-Morley experiment. Nevertheless, the functional behaviour of equation (16) is worth-paying. Consider Figure 4 which is presented to the students after the above discussion:

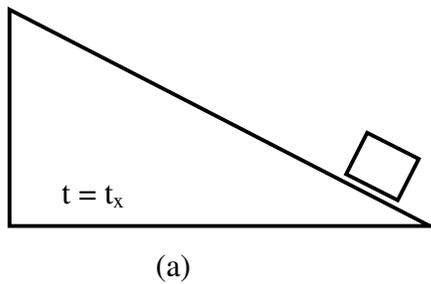
(a)

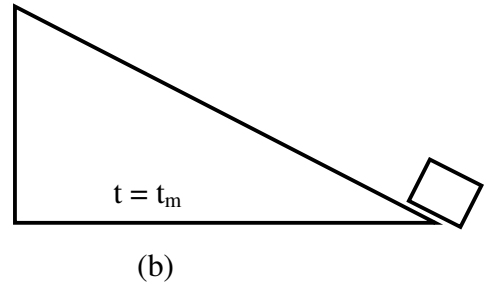
(b)

**Figure 4.** The Lorentz contraction

The argument here can be as follows: suppose that at time $t = t_x$ the position of the brick is that shown in Fig. 4 (a); then we take a photography, which is a kind of measurement. But according to equation (9), that is, according to the *Time Dilation* effect, we will see the brick at the position shown in Fig. 4 (b). In other words, this is what equation (16) tells, that is, the position of the brick which would be recorded by a clock located on and moving along with it is contracted when measured by a chronometer moving with any finite velocity with respect to the body.

Again, this scheme of presentation avoids moving frames; however, most students are able to master the passage from the real experiment to the idealised one of a chronometer moving with respect to the brick., which is equivalent to the methods they will see later in sequential courses of Mechanics.

**CONCLUSIONS**

The strategy above shows that the main topics of Relativity theory can be treated in elementary undergraduate courses with considerable success; according to our experience, this is directly related to the experimental approach plus the element of surprise involved on the determination of the local gravity as opposed to the subsequent description of the motion.

Generally speaking, there is a substantial change of attitude of the students during the process, once they start considering relativity as necessary evolution of newtonian mechanics, substituting their initial curiosity about unusual phenomena by a sounded physical consideration of the consequences of rapid events.

We believe on the possibility of extension of this procedure to secondary school.

## ACKNOWLEDGEMENTS

Useful discussions with L.F. Delboni, I. Matile and A.J.Roberto Jr. are gratefully acknowledged.